\documentclass[12pt]{article}
\usepackage{graphicx}

\newcommand\pubnumber{DPF2013-91}
\newcommand\pubdate{\today}

\def\support{\footnote{Operated by Fermi Research Alliance, LLC under Contract No. De-AC02-07CH11359 with the United States Department of Energy.}}

\def\fermi{
Fermi National Accelerator Laboratory, PO Box 500, Batavia, IL 60510}

\def\Title#1{\begin{center} {\Large #1 } \end{center}}
\def\Author#1{\begin{center}{ \sc #1} \end{center}}
\def\Address#1{\begin{center}{ \it #1} \end{center}}

\newcommand\pubblock{\rightline{\begin{tabular}{l} \pubnumber\\
         \pubdate  \end{tabular}}}
\newenvironment{Abstract}{\begin{quotation}  }{\end{quotation}}
\newenvironment{Presented}{\begin{quotation} \begin{center} 
             PRESENTED AT\end{center}\bigskip 
      \begin{center}\begin{large}}{\end{large}\end{center} \end{quotation}}
\def\Acknowledgments{\bigskip  \bigskip \begin{center} \begin{large}
             \bf ACKNOWLEDGMENTS \end{large}\end{center}}

\def\beq{\begin{equation}}
\def\eeq#1{\label{#1}\end{equation}}
\def\eeqn{\end{equation}}

\def\beqa{\begin{eqnarray}}
\def\eeqa#1{\label{#1}\end{eqnarray}}
\def\eeqan{\end{eqnarray}}

\let\bar=\overbar

\def\Dslash{\not{\hbox{\kern-4pt $D$}}}
\def\dslash{\not{\hbox{\kern-2pt $\del$}}}

\def\msb{{\bar{\ssstyle M \kern -1pt S}}}

\begin{document}
\begin{titlepage}
\pubblock

\vfill
\Title{Intensity Frontier Computing at Fermilab}
\vfill
\Author{ Stephen Wolbers\support}
\Address{\fermi}
\vfill
\begin{Abstract}
The Intensity Frontier (IF) experiments at Fermilab require computing, software, data handling, and 
infrastructure development for detector and beamline design and to extract maximum scientific output
 from the data. The emphasis of computing at Fermilab for many years has been on the Tevatron
  collider Run 2 experiments and CMS. Using the knowledge and experience gained from those
  experiments as well as new computing developments, preparations for computing for IF experiments 
  are ramping up.
There are many challenges in IF computing. These include event generators and detector simulation,
  beamline simulation, detector design and optimization, data acquisition, data handling, data 
  analysis, and all of the associated services required.  In this presentation the computing challenges and requirements will be described and the approaches being taken to address them will be shown.
\end{Abstract}
\vfill
\begin{Presented}
DPF 2013\\
The Meeting of the American Physical Society\\
Division of Particles and Fields\\
Santa Cruz, California, August 13--17, 2013\\
\end{Presented}
\vfill
\end{titlepage}
\def\thefootnote{\fnsymbol{footnote}}
\setcounter{footnote}{0}

\section{Introduction}

Fermilab has a rich and growing suite of Intensity Frontier (IF) experiments.  The Intensity Frontier program is a component of the international and national framework or strategy for particle physics.   Specifically, the Fermilab program includes Intensity, Energy and Cosmic Frontier components.   The Fermilab Intensity Frontier experimental program consists of short baseline and long baseline neutrino experiments, high precision measurements, as well as searches for rare phenomenon.    The MINER$\nu$A, MINOS+, NO$\nu$A and SeaQuest experiments are currently collecting data, MicroBooNE will run in the near future and other experiments such as Muon g-2 and Mu2E as well as LBNE are scheduled to run in the longer term.  In addition a rather extensive test beam program exists.  Many experiments with data continue to analyze data and will do so for quite some time.  Liquid Argon prototype and R$\&$D programs leading to LBNE are also taking data in test setups and test beams. New proposals and R$\&$D, including polarized beams and a neutrino source, are also a component of the Intensity Frontier program. 

\section{Intensity Frontier Computing Requirements}

Computing is essential for the success of the Intensity Frontier program.  In particular it is important to properly recognize and plan the computing so that physics results are not delayed and in fact are maximized.  This was recognized and extensively discussed in the recent Snowmass summer study\cite{snowmass}.  At Fermilab the a transition from Tevatron Run 2 to the Intensity Frontier has led to a large shift of computing resources and effort.

The computing requirements for the Intensity Frontier program are significant.   The aggregate utilization of IF computing at Fermilab approaches the scale of a collider experiment or the CMS Tier 1 center at Fermilab.  This does not include the IF experiments use of grid resources off-site.   The computing requirements are driven by the nature of the experiments, projects and theory in the Intensity Frontier.  

The challenge in IF computing is to respond to the increasing demands driven by new designs, new beam lines, new detectors, and the large-scale analysis of data.  The sophistication of new detectors with fine granularity and excellent particle identification puts large demands on offline simulation and reconstruction programs.  DAQ and trigger system must be able to distinguish between signal and background. The computing systems must be capable of handling all of these new demands.  At the same time experiment services must be migrated to newer services that are more flexible, scalable and aligned with national grid capabilities (Open Science Grid) and with Energy Frontier computing techniques.  All of these must occur in a computing environment that is rapidly changing with more parallel computing, including multi-core systems, GPU's and other new architectures, high network throughput, multi-site computing, virtualization and other technology changes.  

\section{Strategy for Computing at the Intensity Frontier}

Intensity Frontier experiments at Fermilab consist of 30-400 collaborators each.  The experiments are not individually large enough to invent and support the software, data handling and other systems required to collect and analyze data.  Common approaches and solutions are desirable and essential to support this broad range of experiments with the limited effort that is available.  The solutions that are chosen will have to empower the physicists to concentrate on the physics code and physics analysis and not on the software infrastructure around it.   Effort will be made to integrate the core software into robust solutions for each experiment.  Data access and data handling are important aspects of IF computing and access to worldwide resources is an important goal.  Technologies such as multi-core processors, GPU's, High Performance Computing (HPC) and cloud computing all need to be considered.  Training for physicists that allow for contributions from developers, analyzers and others is important.  In this paper I will discuss a few of the common projects and tools that are in development, with an emphasis on projects that involve Fermilab's Scientific Computing Division (SCD).

\begin{figure}[htb]
\centering
\includegraphics[height=2.5in]{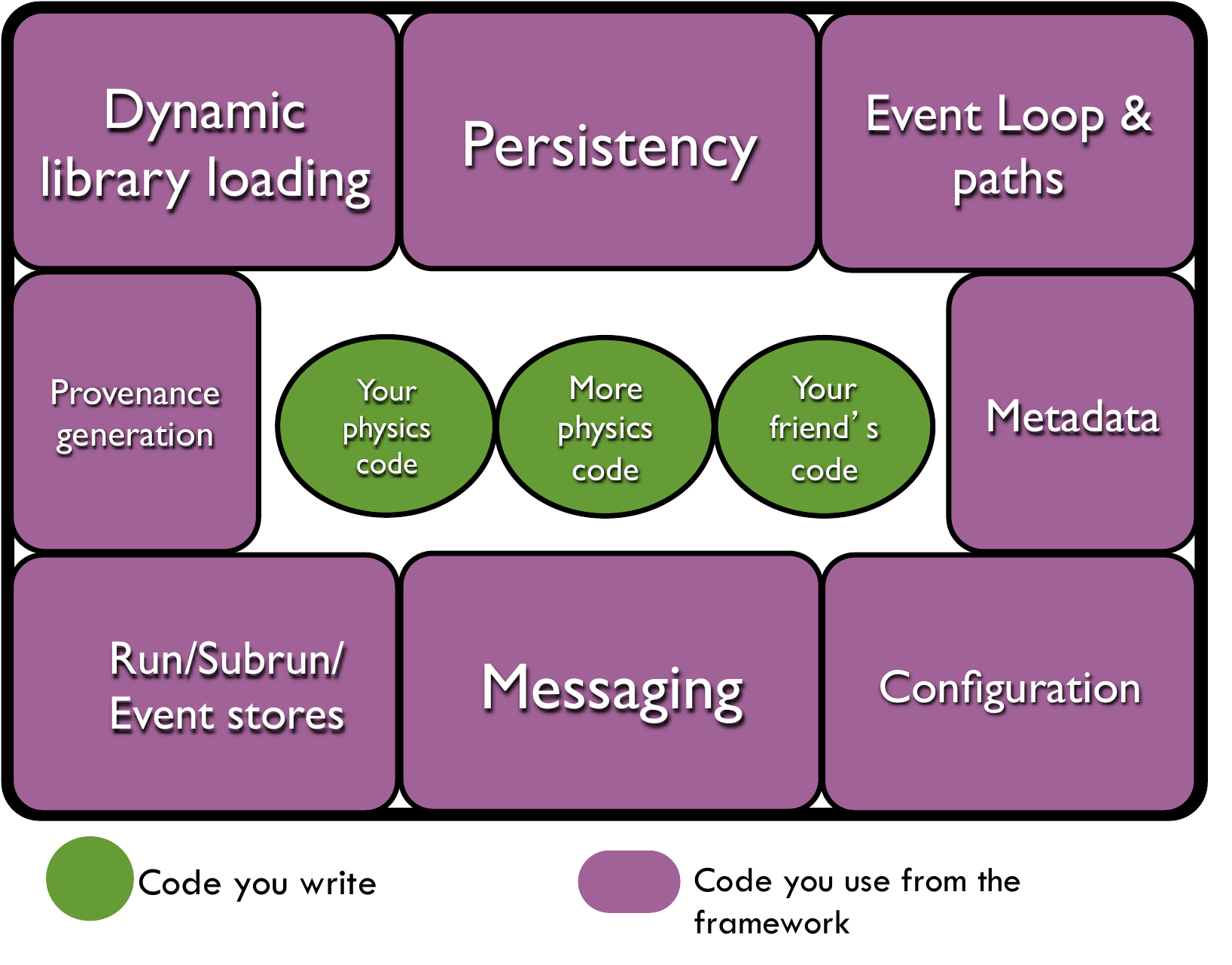}
\caption{Conceptual view of the art framework}
\label{fig:art}
\end{figure}

There are many common projects in various stages of planning and execution.  These include the common framework ``art'', design of an overall computing architecture, generators and simulation codes including GENIE and GEANT4, software toolkits such as LArSoft, many tools and utilities, data handling systems including SAM-web and the FIFE project, computing facilities at Fermilab, and ROOT.  In this paper I will discuss art, LArSoft, the FIFE project, and computing facilities.  

art\cite{art} is a C++ framework developed at Fermilab primarily for the IF experiments.  It is a follow-on of the CMS framework and is currently in use by the Muon g-2, NO$\nu$A, MicroBooNE, LBNE and Mu2e experiments.  This broad use allows for and encourages shared development and support across many experiments and with the developers of the framework.  Integration of Fermilab's data-handling system and other services is an important part of the art project.  A variation for DAQ systems, art-daq, is also in active development and use.  Future directions include hooks for parallel processing at various levels of granularity, including at the sub-event level.  A visual representation of the art framework can be found in Figure~\ref{fig:art}.  The framework handles services and I/O, the user and experiments provide physics code and algorithms.  

LArSoft\cite{LArSoft} is a common simulation, reconstruction and analysis toolkit for experiments that use liquid argon time projection chambers.  The package is managed by the Fermilab Scientific Computing Division.  SCD's emphasis is on code packaging and management, release procedures, integration and coordination with the art framework, and close cooperation with the experiments that use the package and provide algorithm development effort.  Again the common effort allows multiple experiments to benefit and progress from this shared activity.  ArgoNeut, MicroBooNE, LBNE, LArIAT and various liquid argon R$\&$D activities are all using LArSoft. 

The next common project to discuss is FIFE\cite{FIFE}, FabrIc for Frontier Experiments.  FIFE is a Fermilab led project that aims to provide an integrated framework for offline analysis.  It is focused on Intensity Frontier experiments but is not exclusively designed for them.  Relevant aspects of the FIFE project include providing a complete system for computing, modularity to allow experiments to use some but not all of the features if desired, excellent design, collaborative work with experimenters, ability to use distributed resources, and the ability to use tools from outside the Fermilab and HEP community.  The goal of the project is to allow the experiments, Fermilab and any outside institutions to work together to enable maximal physics results at a minimum cost.

\begin{figure}[htb]
\centering
\includegraphics[height=3.5 in]{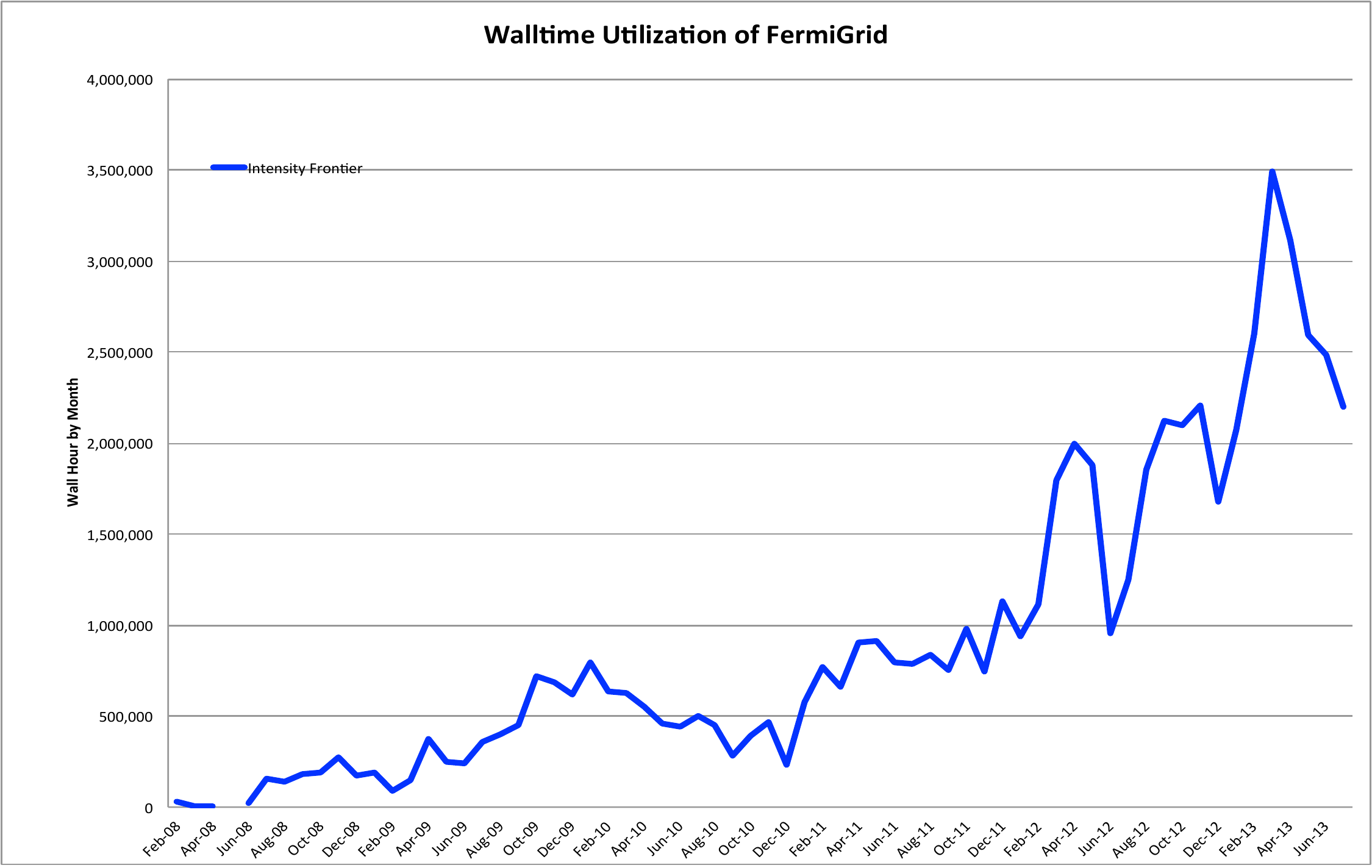}
\caption{Recent Growth of Intensity Frontier CPU use (wall clock hours/month) at Fermilab}
\label{fig:CPU}
\end{figure}

Computing facilities at Fermilab and elsewhere are required to provide computing resources needed by the Intensity Frontier program.  At Fermilab a shared services model has been used productively for many years.  This allows experiments to use services throughout their lifetime, from early R$\&$D to proposal to project and into data taking and to final analysis and data preservation.   There are many aspects of computing that the IF program requires, including CPU, storage, networking, and all of the tools and services required for collaborations to function.  Figure~\ref{fig:CPU} shows the rapid increase of the CPU used by the Intensity Frontier at Fermilab.  Individual experiments routinely use more than 1000 cores and this is expected to increase dramatically as experiments collect additional data, the intensity of the beams increases, and simulations for future experiments ramp up.  

\begin{figure}[htb]
\centering
\includegraphics[height=3.0in]{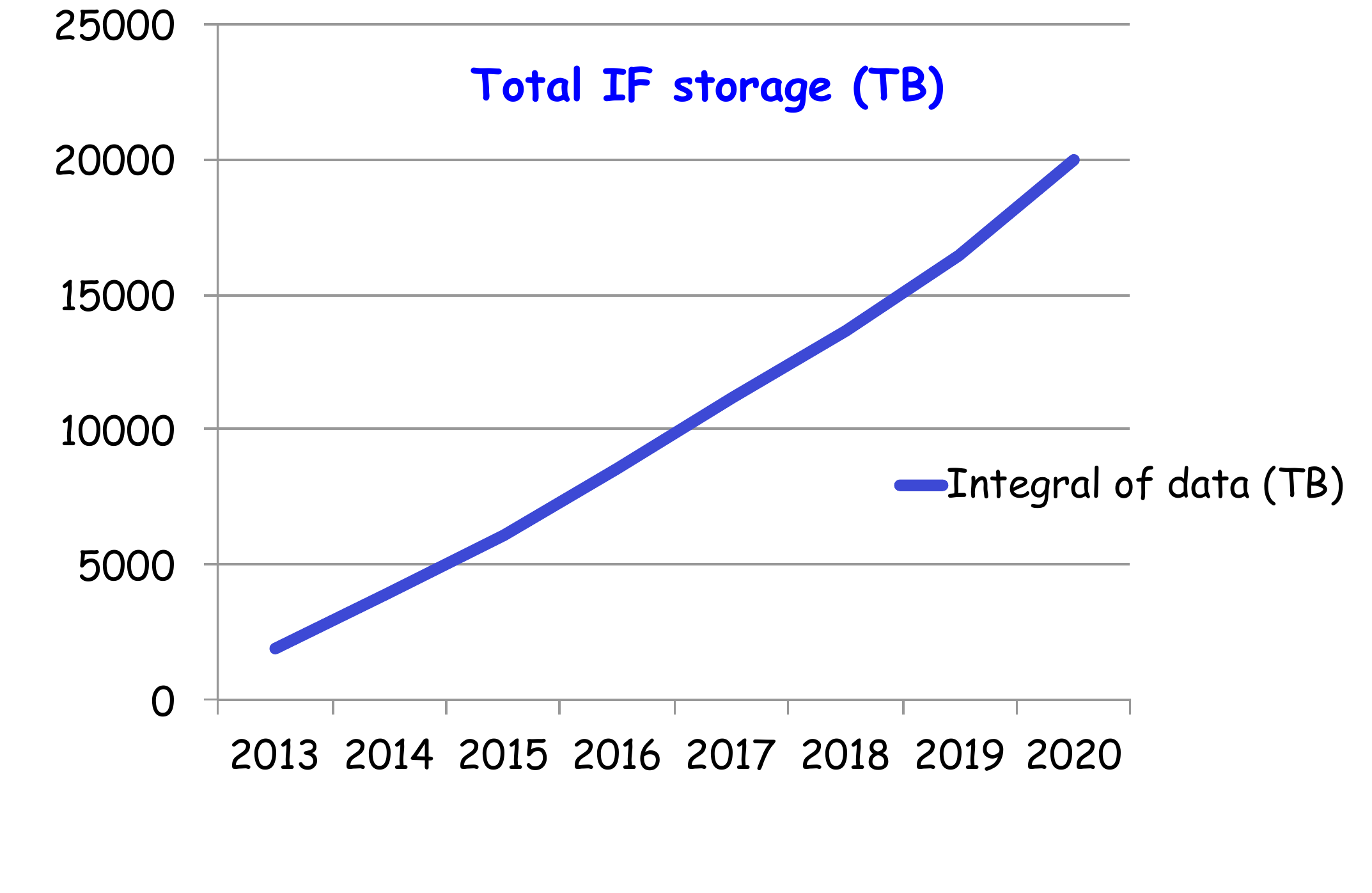}
\caption{Projected growth of Intensity Frontier storage use at Fermilab}
\label{fig:Storage}
\end{figure}

The storage requirements for the Intensity Frontier experiments have been small compared to Energy Frontier experiments.  Currently the Intensity Frontier program uses about 2 PByte on the Fermilab mass storage systems (tape-based archive), compared to about 10 PByte each for CDF and D0 and about twice that for CMS.  However, IF storage has increased and is expected to increase rapidly through 2020.  This was studied at the recent Snowmass workshop.  An early estimate of the predicted ramp up of IF storage through 2020 is shown in Figure~\ref{fig:Storage}.

Fermilab's computing facilities are sufficient to accommodate the current and expected growth in computing required by the Intensity Frontier program.  This includes the computer rooms and the infrastructure associated with them (space, power and cooling), Fermilab internal networking  and external networking to connect to the Open Science Grid and other partners, CPU, disk storage and tape storage.  Upgrades to the computer rooms are in progress and will allow additional room for growth.   To ensure that the experiments and the laboratory are aligned in the mix of resources available and the relative priorities to be placed on hardware purchases and effort, as well as what emphasis is to be placed on various aspects of computing, a yearly scientific portfolio review has been initiated.  The next review should occur early in fiscal year 2014 to help inform and prioritize major purchases and effort as the Intensity Frontier experiments collect and analyze data from the current and previous accelerator runs and as future experiments continue to ramp up.

\section{Conclusion}

Fermilab has recently increased its emphasis on the Intensity Frontier program and continues to transition from Tevatron Run 2 to Intensity Frontier experiments.  Computing is essential to the success  of the Intensity Frontier program.  Facilities, effort and projects aimed at improving the computing capabilities and usability are ramping up and will continue as the Intensity Frontier program grows and matures.  The emphasis is on common and shared tools and solutions, connected with national and international efforts.  The goal is to ensure smooth and effective physics results from the experiments as quickly as possible.

\Acknowledgments
I am grateful to Adam Lyon, Mike Kirby, Ruth Pordes, Margaret Votava, Andrew Norman for their help in preparing the talk.

\end{document}